\documentclass[aps,prb,twocolumn,floatfix]{revtex4-2}

\usepackage{mathtools}
\usepackage{amssymb,amsmath}
\usepackage[dvipsnames]{xcolor}
\usepackage{graphicx}
\usepackage{soul}
\usepackage[colorlinks=true,linkcolor=blue,citecolor=blue,urlcolor=blue]{hyperref}

\newcommand{\vect}[1]{\boldsymbol{#1}}
\newcommand{\AppRef}[1]{Appendix~\hyperref[app:#1]{#1}}

\begin{document}

\title{Extreme-Value Criticality and Gain Decomposition at the Integer Quantum Hall Transition}

\author{Wei-Han Li}\email{wei-han.li@outlook.com}

\author{Abbas Ali Saberi}
\email{asaberi@constructor.university}
\affiliation{School of Science, Constructor University, Campus Ring 1, 28759 Bremen, Germany \\ Max Planck Institute for the Physics of Complex Systems, 01187 Dresden, Germany}

\begin{abstract}
Extreme-value fluctuations at quantum critical points remain poorly understood in the presence of strong correlations and openness. At the integer quantum Hall transition in the open Chalker--Coddington network, we show that the maximal wave-function amplitude separates into a global gain and an intrinsic extreme component, $|\psi|_{\max}=A\,|\tilde{\psi}|_{\max}$. We introduce extreme-moment scaling for $|\psi|_{\max}$ and observe an approximately parabolic exponent function $\tau_{\max}(q)$ over moderate $q$, while $\ln|\psi|_{\max}$ displays an almost Gaussian bulk over the studied sizes. The gain factor is close to log-normal and largely controls the raw extremes. Gain normalization reorganizes the statistics: $\tilde{\tau}_{\max}(q)$ changes qualitatively and $|\tilde{\psi}|_{\max}$ does not support a single-parameter generalized extreme-value collapse under standard centering/scaling in the accessible size window. Extreme observables thus provide a robust probe of correlated criticality in open quantum systems.
\end{abstract}
\maketitle

\textit{Introduction.}---Extreme fluctuations often expose physics that remains hidden at the level of typical observables. In disordered and random systems, extreme-value (EV) statistics have become an important tool for characterizing rare events, from low-energy excitations and trapping phenomena to fluctuating interfaces and random-matrix spectra \cite{Bou1997, Dou2003, Fyo2008, Ray2001, Gyo2003, Ram2010, Meh2004}. This perspective is especially useful in strongly correlated systems, because the standard EV universality classes are derived for independent or weakly dependent variables, so systematic departures from them can directly encode correlation effects \cite{Fisher1928, Gnedenko1943, Leadbetter1983}. Yet, despite the broad development of EV methods in statistical physics \cite{Igl2020, Her2021}, the EV statistics of quantum eigenstates in strongly correlated or critical settings remain less explored, with only a few studies in special ensembles and related models \cite{Lak2008, Li2024}.

Extreme-value statistics of multifractal patterns have been explored in broader contexts such as logarithmically correlated fields and the associated “freezing” scenario, where departures from i.i.d.\ generalized extreme-value universality can be intrinsic and convergence may be anomalously slow, with logarithmic corrections \cite{FyoGiraud2015}. This motivates developing extreme-value observables for quantum-critical states, where multifractal correlations can strongly reshape the statistics of maxima and, in open or driven geometries, additional sample-dependent global amplification/normalization factors may arise that must be separated from intrinsic extremal fluctuations.

A paradigmatic arena in which critical eigenstates are the central objects is the integer quantum Hall (IQH) plateau transition, which has also been approached from complementary spectral and network-based perspectives \cite{KettemannTsvelik1999}. At criticality, electronic wave functions are delocalized and exhibit multifractal fluctuations, a defining feature of localization--delocalization criticality in two dimensions. Over the past decades, multifractal analysis has been developed extensively for the IQH problem, both numerically and analytically, and has revealed highly nontrivial scaling structure beyond simple single-parameter descriptions \cite{Eve2001,Eve2008,Kramer2005,Obu2008,Obu2010,Hua2021,Bab2023}. Most of this literature, however, focuses on bulk moment observables (inverse participation ratios and related quantities), which characterize amplitude statistics through ensemble moments and their scaling. A complementary question remains largely open for IQH critical wave functions: what is the scaling theory of the most extreme wave-function amplitudes? This question is physically natural, in principle experimentally relevant, and conceptually nontrivial because maxima are governed by the interplay of rare events, scale-invariant critical correlations, and finite-size effects.

In this Letter, we study extreme wave-function amplitudes at the integer quantum Hall transition in the Chalker--Coddington (CC) network \cite{Cha1988} (see \cite{Kramer2005} for broader context), focusing on the open, point-contact geometry \cite{Bon2014}. We find that the maximum in the driven stationary state is \emph{not} a purely intrinsic critical observable: it factorizes into a sample-dependent global gain and an intrinsic extremal component, $|\psi|_{\max}=A\,|\tilde\psi|_{\max}$, a structure that is absent in conventional analyses of self-normalized IQH eigenstates and, to our knowledge, has not been identified for extreme observables in this setting. This gain decomposition is the central new element of our work and complements known exact relations linking wave-function multifractality to the multifractality of Wigner delay times in lead-attached critical systems \cite{Mir2006}.

We formulate an extreme-value multifractality framework for the open CC model. We define extreme-moment scaling for the maximum and extract an exponent function $\tau_{\max}(q)$ and the associated large-deviation spectrum for maxima. Over the accessible moment window, the raw extrema exhibit an approximately parabolic $\tau_{\max}(q)$ and $\ln|\psi|_{\max}$ is close to Gaussian, in marked contrast to standard EV expectations for weakly dependent variables \cite{Fisher1928,Gnedenko1943,Leadbetter1983}. We show that these raw extreme statistics are strongly shaped by the gain factor $A$, whose fluctuations are close to log-normal, while gain normalization reorganizes both the exponent function and the probability density, isolating an intrinsic, correlation-dominated extreme sector beyond collective amplification.

\begin{figure}[t]
\includegraphics[width=\columnwidth]{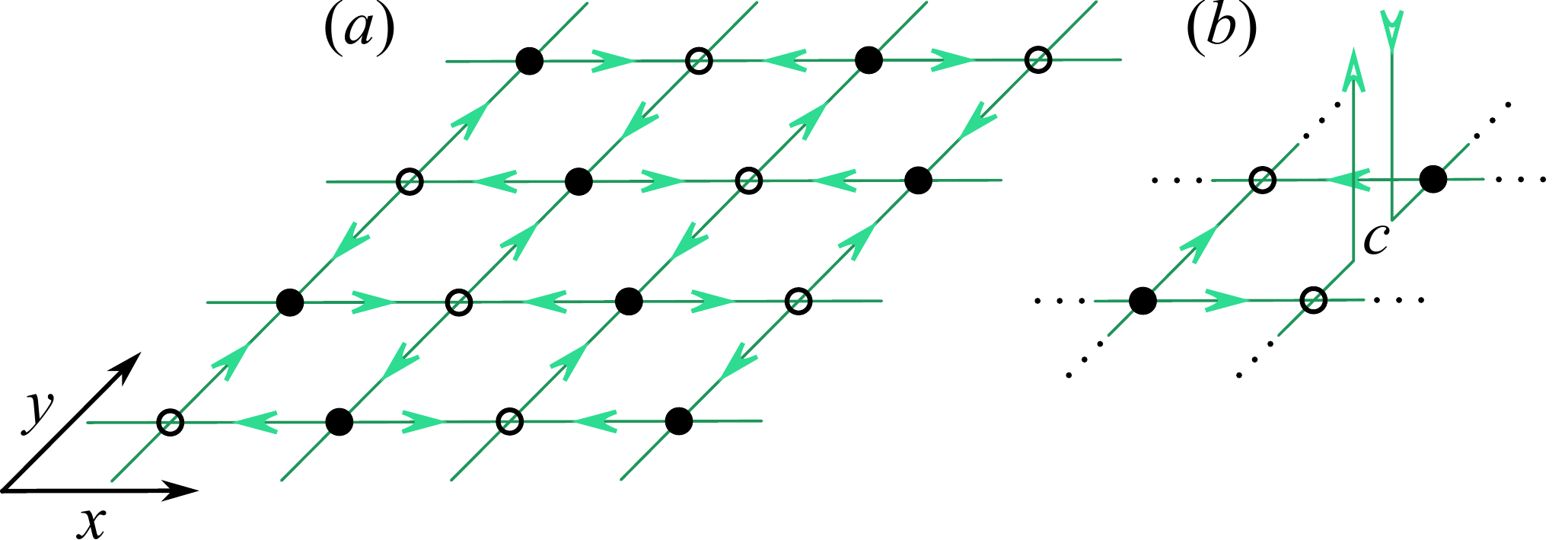}
\centering
\caption{Chalker--Coddington (CC) network on a 2D square lattice. (\textit{a}) The wave function lives on directed links and propagates along arrows, forming closed loops with alternating circulation on neighboring plaquettes. At each node, a local $2\times2$ scattering matrix mixes two incoming into two outgoing amplitudes; $\bullet$ and $\circ$ indicate the two node orientations. (\textit{b}) Open CC network with an external point contact attached to link $c$.}
\label{fig_1}
\end{figure}

\textit{CC networks.}---
We study the Chalker--Coddington network model on a square lattice, shown in Fig.~\ref{fig_1}(a), which provides a standard lattice description of the IQH plateau transition \cite{Cha1988, Kramer2005}. The wave function lives on directed links
$|\varPsi\rangle=\sum_l \psi_l |l\rangle$ with $\psi_l\in\mathbb{C}$,
and evolves in discrete time under a unitary operator $\hat U$ built from local scattering matrices at network nodes (see \AppRef{A} for the convention for $\hat U$ and the open-network stationary state).
At the critical point, the scattering probabilities are tuned to be isotropic, and the network lies in the IQH universality class.

We focus on the open CC network \cite{Bon2014}, implemented by attaching an external contact link $|c\rangle$ [Fig.~\ref{fig_1}(b)]. The stationary state is obtained from
\begin{equation}\label{states}
(1-\mathcal{Q}\hat U)|\varPsi\rangle=|c\rangle,
\qquad
\mathcal{Q}=1-|c\rangle\langle c|,
\end{equation}
which defines a unique driven steady state in the open geometry  (see \AppRef{A}).

\textit{Extreme-value multifractality.}---
We analyze the stationary solution of Eq.~(\ref{states}) through the maximum link amplitude
\begin{equation}
|\psi|_{\max} \equiv \max\{|\psi_{l\neq c}|\},
\end{equation}
excluding the contact link since $|\psi_c|=1$ is fixed by the injection convention of the open dynamics.

To characterize its finite-size scaling, we introduce the extreme moments
\begin{equation}\label{tau_q}
\mathbb{E}\!\left[(|\psi|_{\max})^{2q}\right]\sim L^{-d\,\tau_{\max}(q)},
\end{equation}
where $\mathbb{E}\{\cdots\}$ denotes disorder averaging and $d=2$ is kept explicit as our scaling convention. By definition, the $q=0$ moment equals unity, hence $\tau_{\max}(0)=0$.

The open geometry motivates two generic expectations, both confirmed numerically: (i) $|\psi|_{\max}$ is typically larger than unity, and (ii) its typical scale increases with system size. Consequently, $\mathbb{E}[(|\psi|_{\max})^{2q}]$ increases with $L$ for $q>0$ and decreases with $L$ for $q<0$, implying $\tau_{\max}(q>0)<0$ and $\tau_{\max}(q<0)>0$. This sign structure already distinguishes the EV sector of the open network from conventional multifractal moments of normalized wave functions. 

\textit{Parabolic extreme-moment exponent.}---
Our first central result is that, in the accessible moment window $|q|\lesssim 1$, the extreme-moment exponent is accurately described by a parabolic form,
\begin{equation}\label{parabolic}
\tau_{\max}(q)\approx -\gamma q^2+\nu q,
\end{equation}
with $\nu=-0.430\pm0.002$ and $\gamma=0.137\pm0.0007$, obtained from finite-size scaling (see \AppRef{C}). 

To expose the quadratic structure more transparently, we define the shifted extreme moments
\begin{equation}
\mathbb{E}\!\left[p^{\,2q}\right]\sim L^{-d\,\Delta_q^{\max}}
\end{equation}
with $p \coloneqq L^{\nu d/2}|\psi|_{\max}$, so that Eq.~(\ref{parabolic}) implies
\begin{equation}\label{Delta_q}
\Delta_q^{\max}\approx -\gamma q^2 .
\end{equation}

\begin{figure}[t]
\includegraphics[width=\columnwidth]{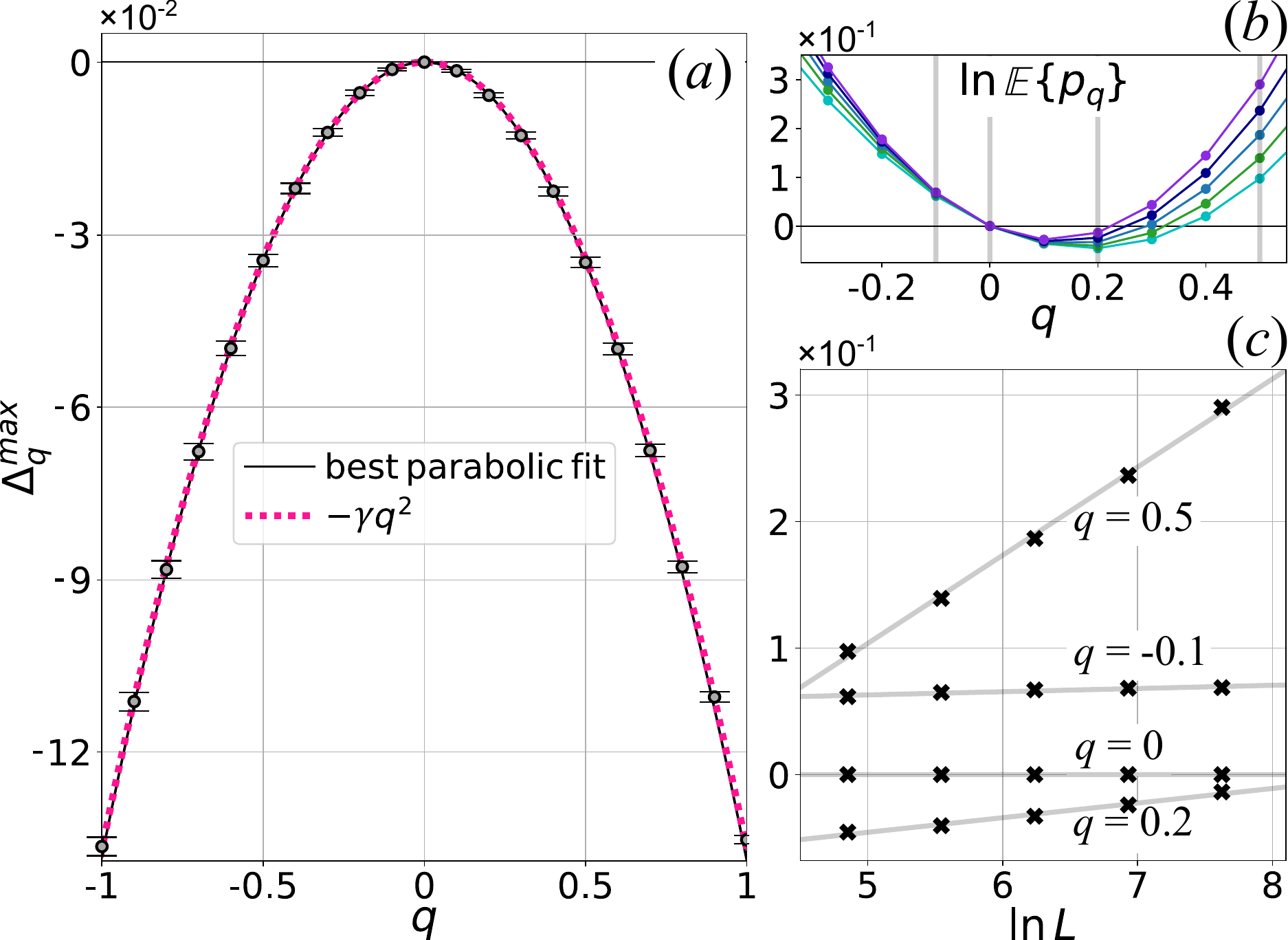}
\centering
\caption{Parabolic extreme-moment exponent. (a) Measured $\Delta_q^{\max}$ for $|q|\lesssim 1$. The dashed curve shows the prediction $\Delta_q^{\max}\approx -\gamma q^2$, while the solid curve is an unconstrained best parabolic fit. (b) $\ln \mathbb{E}\!\left[p^{\,2q}\right]$ versus $\ln L$ (equivalently, the moments of the shifted variable entering the fit). Data from bottom to top correspond to $L=64,128,\ldots,2048$; ensemble sizes are of order $\mathcal{O}(10^7)$ for $L<1000$ and $\mathcal{O}(10^6)$ for $L>1000$. (c) Representative extractions of $\Delta_q^{\max}$ from the slopes of linear fits of $\ln \mathbb{E}\!\left[p^{\,2q}\right]$ versus $\ln L$.}
\label{fig_2}
\end{figure}

Figure~\ref{fig_2} shows that $\Delta_q^{\max}$ follows the predicted parabola with excellent accuracy throughout the measured range. The agreement is quantified by adjusted $\bar R^2>0.999$ for the parabolic fits (see \AppRef{B}). At the same time, we emphasize that this parabolicity is established only over a finite $q$ interval; whether it persists asymptotically at larger $|q|$ remains an open question.

Importantly, the parabolic form in Eq.~(\ref{parabolic}) is \emph{distinct} from the conventional multifractal exponents of the full (all-link) wave-function statistics (see \AppRef{C} and \AppRef{D}), underscoring that the maximum probes a different rare-event sector. At the same time, this parabolicity pertains to the \emph{raw} maximum in the open geometry and therefore incorporates the effect of a fluctuating global gain field. As we show below, gain normalization reorganizes the extreme-moment exponent and drives systematic deviations from a simple parabola, demonstrating that the raw parabolicity is not purely intrinsic to the extremal sector.

\textit{Parabolic extreme-value singularity spectrum}---
To complement the extreme-moment exponents, we characterize sample-to-sample fluctuations of the maximum by a large-deviation description on the scale $\ln L$. For each realization, define $\alpha$ by $|\psi|_{\max}^2 \coloneqq L^{-d\,\alpha}$,
equivalently
\begin{equation}\label{alpha}
\alpha=-\frac{\ln |\psi|_{\max}^2}{d\,\ln L},
\end{equation}
which need not be positive since the open-network gain can produce $|\psi|_{\max}>1$. We introduce an extreme-value singularity spectrum $f_{\max}(\alpha)$ through
\begin{equation}\label{LD_alpha}
\mathrm{PDF}_L(\alpha)\asymp L^{d\,f_{\max}(\alpha)}\qquad (L\to\infty),
\end{equation}
where $\asymp$ denotes equality at the level of the leading exponent in $\ln L$ \cite{Halsey1986,Touchette2009}. Normalization implies $\max_\alpha f_{\max}(\alpha)=0$.

The connection to $\tau_{\max}(q)$ follows from Laplace (saddle-point) evaluation of
\begin{equation}\label{moment_alpha_int}
\begin{aligned}
\mathbb{E}\!\left[(|\psi|_{\max})^{2q}\right]
&=\int \mathrm{d}\alpha\,\mathrm{PDF}_L(\alpha)\,L^{-d q\alpha}\\
&\asymp \int \mathrm{d}\alpha\,
\exp\!\Big\{d\,\ln L\big[f_{\max}(\alpha)-q\alpha\big]\Big\}.
\end{aligned}
\end{equation}
Comparing with $\mathbb{E}[(|\psi|_{\max})^{2q}]\sim L^{-d\,\tau_{\max}(q)}$ yields the usual Legendre relations
\begin{equation}\label{legendre}
f_{\max}(\alpha)=q\alpha-\tau_{\max}(q),\qquad
\alpha=\partial_q\tau_{\max}(q),
\end{equation}
with saddle condition $q=\partial_\alpha f_{\max}(\alpha)$. Here $f_{\max}(\alpha)$ describes large deviations of the \emph{single} maximum per sample (an extreme-value spectrum), not the conventional multifractal spectrum that encodes the geometric distribution of intensities over all links.

Using the parabolic form in Eq.~(\ref{parabolic}), we obtain
\begin{equation}\label{fmax_parabola}
f_{\max}(\alpha)\approx -\frac{(\alpha-\nu)^2}{4\gamma}
\end{equation}
(see \AppRef{C}). Thus, in the measured regime $|q|\lesssim 1$, the raw maxima are governed by a parabolic spectrum. A quadratic $f_{\max}(\alpha)$ implies that $\mathrm{PDF}_L(\alpha)$ is approximately Gaussian near its typical value on the $\ln L$ scale, and therefore $\ln|\psi|_{\max}$ has an approximately Gaussian bulk (equivalently, near-log-normal raw maxima) over the same moment window. This is nontrivial: for weakly dependent variables, one expects convergence to the standard GEV classes rather than near-log-normal behavior \cite{Fisher1928,Gnedenko1943,Leadbetter1983}. In the present open geometry, the observed near-Gaussian bulk is further shaped by the fluctuating gain discussed below.

\begin{figure}[t]
\includegraphics[width=\columnwidth]{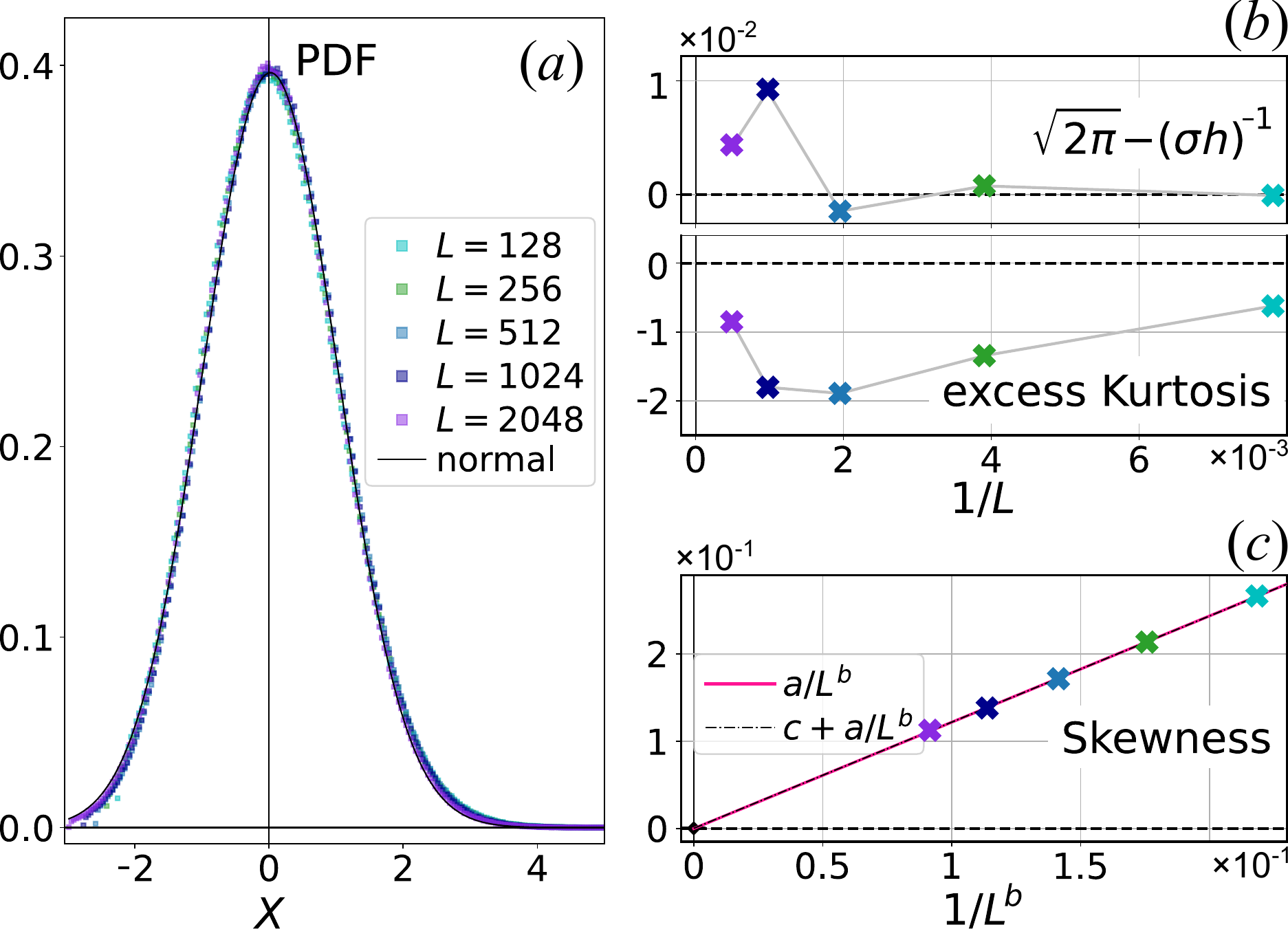}
\centering
\caption{Near-log-normal statistics of the raw maxima. (a) Probability densities of $\ln |\psi|_{\max}$ shown in terms of the standardized variable $X=(\ln |\psi|_{\max}-\mu)/\sigma$, where $\mu$ and $\sigma$ are the mode and standard deviation of $\mathrm{PDF}(\ln |\psi|_{\max})$, respectively. The curves approach a normal distribution. (b) Two diagnostics of Gaussianity, $\sqrt{2\pi}-(\sigma h)^{-1}$ and the excess kurtosis (with $h$ the peak height of the PDF), remain close to zero; the larger scatter at the largest $L$ is due to finite ensemble size. (c) Skewness versus $L$. The data are well fitted by $aL^{-b}$ (solid line) with $b=0.314\pm0.002$ and $a=1.218\pm0.016$, while the more general form $aL^{-b}+c$ gives $c\approx 3\times 10^{-4}$ (dash-dotted line), consistent with vanishing skewness as $L\to\infty$.}
\label{fig_3}
\end{figure}

Figure~\ref{fig_3} demonstrates this behavior directly. After centering and scaling, $\mathrm{PDF}(\ln |\psi|_{\max})$ collapses toward a normal curve. We quantify the approach to Gaussianity using the skewness, excess kurtosis, and $\sqrt{2\pi}-(\sigma h)^{-1}$ (with $h$ the peak height of the PDF), which vanish for an exact Gaussian. The excess kurtosis and height-based diagnostic remain close to zero across sizes, with mild deviations at the largest $L$ attributable to finite sampling. The skewness decays as $aL^{-b}$ with $\bar R^2>0.9996$ (see \AppRef{B}), providing strong evidence that the standardized bulk of $\mathrm{PDF}(\ln |\psi|_{\max})$ approaches a Gaussian form as $L$ increases (within the sampled window); tail asymptotics may involve additional finite-size or correlation-induced structure. In our largest systems ($L=1024,2048$), the ensemble size is of order $\mathcal{O}(10^6)$, which sets the scale of the residual statistical noise.

\begin{figure}[t]
\includegraphics[width=\columnwidth]{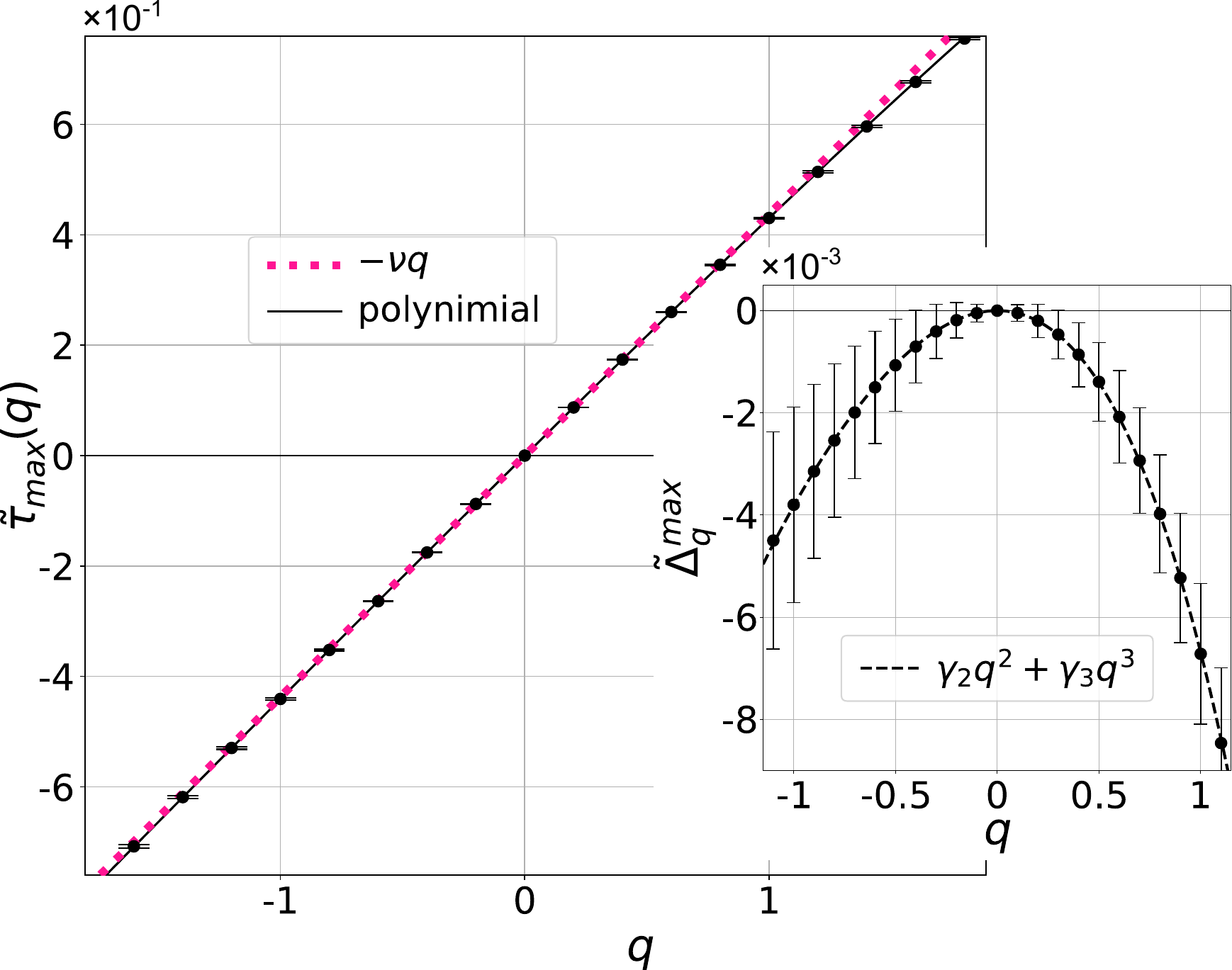}
\centering
\caption{Scaling exponents for gain-normalized maxima $|\tilde\psi|_{\max}$. The exponent $\tilde\tau_{\max}(q)$ (solid line) is dominated by a linear contribution $\approx -\nu q$ (dashed line). Inset: the nonlinear remainder $\tilde\Delta_q^{\max}$ shows a small but systematic parabolic term and a cubic correction, generating an asymmetry between $q$ and $-q$. The fitted coefficients are $\gamma_2=-0.005\pm2.54\times10^{-5}$ and $\gamma_3=-0.0014\pm8.06\times10^{-6}$; higher-order terms are also present but subleading in the plotted range.}
\label{fig_4}
\end{figure}

\textit{Gain normalization and intrinsic extreme sector.}---
The central physical ingredient of the open network is the sample-dependent global gain
$A=\left(\sum_{l\neq c} |\psi_l|^2\right)^{1/2}$,
which rescales link amplitudes collectively within each realization and fluctuates from sample to sample (its statistics are provided in \AppRef{E}). The quantity $A^{2}=\sum_{l\neq c}|\psi_l|^{2}$ is the total stationary-state intensity (excluding the fixed contact amplitude). In lead-attached critical systems, closely related integrated intensities are tied to scattering observables such as Wigner delay times and their moments \cite{Mir2006}; our “gain” factor isolates the resulting global amplification component of the extremal statistics.

This global factor contributes directly to the raw maximum and strongly shapes the observed scaling and near-log-normality. To isolate the intrinsic extremal fluctuations, we therefore define the gain-normalized maximum
\begin{equation}
|\tilde\psi|_{\max}\coloneqq \frac{|\psi|_{\max}}{A}.
\end{equation}

The normalized extremes obey their own scaling law, $\mathbb{E}\!\left[|\tilde\psi|_{\max}^{2q}\right]\sim L^{-d\,\tilde\tau_{\max}(q)}$,
with the resulting exponent function qualitatively different from the raw one. As shown in Fig.~\ref{fig_4}, $\tilde\tau_{\max}(q)$ is dominated by a linear term, well approximated by $-\nu q$, while the residual nonlinear part
$\tilde\Delta_q^{\max}\equiv \tilde\tau_{\max}(q)+\nu q$
is small and asymmetric in $q$, with visible quadratic and cubic contributions. This breakdown of the simple parabolic form demonstrates that the parabolicity of the raw maximum is not an intrinsic property of the extremal fluctuations alone; rather, it emerges from the interplay between critical extremes and the global gain field in the open geometry.

The probability density of the normalized maxima corroborates this picture. In contrast to the raw variable, $\mathrm{PDF}(\ln |\tilde\psi|_{\max})$ is no longer close to a simple log-normal form. Instead, it exhibits a compound structure with a Gaussian tail at large values together with systematic deviations from a single-parameter GEV description under the same centering/scaling protocol across sizes (see \AppRef{E}). This supports that the normalized maximum defines a distinct intrinsic extreme sector, consistent with correlation-dominated critical fluctuations.

\textit{Conclusions.}---
We have introduced an extreme-value scaling framework for critical wave functions at the integer quantum Hall transition and applied it to the open Chalker--Coddington network. The raw maximum $|\psi|_{\max}$ exhibits an extreme-moment exponent $\tau_{\max}(q)$ that is approximately parabolic in the accessible range and an extreme singularity spectrum with the corresponding quadratic form. Consistently, the standardized bulk of $\ln |\psi|_{\max}$ is well described by a Gaussian distribution over accessible sizes, i.e., the raw maxima are near-log-normal in this regime rather than governed by standard i.i.d.\ extreme-value theory.

The key physical result is that the open geometry endows the stationary state with a fluctuating global gain factor $A$, so the raw maximum combines collective amplification and intrinsic extremal fluctuations. After removing this gain, the normalized maximum $|\tilde\psi|_{\max}$ displays a qualitatively different exponent function and a nontrivial probability density, revealing a distinct intrinsic extreme sector. This multiplicative decomposition of critical extremes is specific to open critical systems and lies beyond the scope of conventional multifractal analyses of self-normalized wave functions.

Our results establish extreme observables as a sensitive probe of correlated quantum criticality in open systems and open a route to studying rare-event sectors at localization transitions beyond standard moment-based multifractality.

\textit{Acknowledgements.}---This work was funded by the Deutsche Forschungsgemeinschaft (DFG, German Research Foundation) under Project No.~557852701 (A.A.S.). The study was also supported by the Advanced Study Group ``Strongly Correlated Extreme Fluctuations'' at the Max Planck Institute for the Physics of Complex Systems, Dresden (2024/25) \cite{pks_asg2024}.

\bibliographystyle{apsrev4-2}
\bibliography{Ref}

\appendix

\section*{Appendix}

This Appendix provides the technical details of the open Chalker--Coddington (CC) network, the definition of the adjusted $\bar{R}^2$ used throughout the paper, benchmark full-state multifractal exponents, the parabolic scaling analysis of the raw maxima $|\psi|_{\max}$, and the statistics of the global gain and normalized maxima $|\tilde\psi|_{\max}$.

\section{Appendix A: Dynamics and stationary states of open CC networks}
\label{app:A}

\renewcommand{\thefigure}{A\arabic{figure}}
\setcounter{figure}{0}
\renewcommand{\theequation}{A\arabic{equation}}
\setcounter{equation}{0}

We use the standard Chalker--Coddington (CC) network at criticality; definitions and conventions for the closed network and its unitary one-step evolution operator $\hat U$ are given in
Refs.~\cite{Cha1988,Eve2001} (see also the review \cite{Eve2008}).
The ``open'' point-contact formulation we employ follows the scattering/absorption setup of
Ref.~\cite{Bon2014}; here we only record the minimal equations needed to define the stationary state used in the main text.

\paragraph{One-step evolution.}
On the directed-link basis $\{|l\rangle\}$, the CC step is $\hat U=\hat P\,\hat S$, where $\hat S=\bigoplus_{\vect{j}}\hat S_{\vect{j}}$ applies the $2\times2$ node scatterers in parallel and $\hat P$ propagates outgoing link amplitudes to the incoming links of the next step \cite{Cha1988,Eve2001}.
At criticality $t=r=1/\sqrt{2}$ (random phases may be placed on nodes or links by a gauge choice \cite{Eve2001,Eve2008}).

We designate a single directed link $c$ as the point contact and introduce $\mathcal Q = 1-|c\rangle\langle c|$.
The open dynamics (injection at $c$ with absorption upon return) is defined by
$|\varPsi\rangle_{t+1}=\mathcal Q \hat U\,|\varPsi\rangle_t + |c\rangle$,which fixes the contact amplitude by construction, $\langle c|\varPsi\rangle_{t}=1$ for all $t$.
A stationary state $|\varPsi\rangle$ satisfies $|\varPsi\rangle_{t+1}=|\varPsi\rangle_t$, hence
$(1-\mathcal Q \hat U)\,|\varPsi\rangle = |c\rangle$,
which is the linear system used in the main text.

In the open geometry, probability leaks into the absorbing contact, so $\mathcal Q\hat U$ is strictly subunitary in the generic disordered case \cite{Bon2014}, implying $\rho(\mathcal Q\hat U)<1$ and thus invertibility of $1-\mathcal Q\hat U$.
Equivalently,
$|\varPsi\rangle=(1-\mathcal Q\hat U)^{-1}|c\rangle=\sum_{n\ge 0}(\mathcal Q\hat U)^n|c\rangle$,
with the usual interpretation as a sum over paths launched from the contact and terminated upon hitting it again \cite{Bon2014}.

\section{Appendix B: The $\bar{R}^2$ value}
\label{app:B}

\renewcommand{\thefigure}{B\arabic{figure}}
\setcounter{figure}{0}
\renewcommand{\theequation}{B\arabic{equation}}
\setcounter{equation}{0}

This section defines the adjusted $\bar{R}^2$ used in the main text to quantify fit quality.
For a data set $X$ and the corresponding fitted values $X_{\mathrm{fit}}$, we use
\begin{equation}
\bar{R}^2(X)
=
1-\frac{(n-1)\,V(X,X_{\mathrm{fit}})}{(n-m-1)\,V_{\mathrm{sd}}(X)} ,
\end{equation}
where $n$ is the number of data points (here, the number of system sizes $L$), and $m$ is the number of fitting parameters.
The variance of the data is
$V_{\mathrm{sd}}(X)=\mathbb{E}\{(X-\bar{X})^2\}$,
and the residual variance is
$V(X,X_{\mathrm{fit}})=\mathbb{E}\{(X-X_{\mathrm{fit}})^2\}$.
The ratio $V(X,X_{\mathrm{fit}})/V_{\mathrm{sd}}(X)$ measures the fraction of variance not captured by the fit. Accordingly, $\bar{R}^2\in[0,1]$ quantifies the explained variability while penalizing over-parameterized fits, and is therefore suitable for comparing nested fitting forms.

\section{Appendix C: Parabolic scaling of maxima}
\label{app:D}

\renewcommand{\thefigure}{C\arabic{figure}}
\setcounter{figure}{0}
\renewcommand{\theequation}{C\arabic{equation}}
\setcounter{equation}{0}

\begin{figure}[t]
\includegraphics[width=0.85\columnwidth]{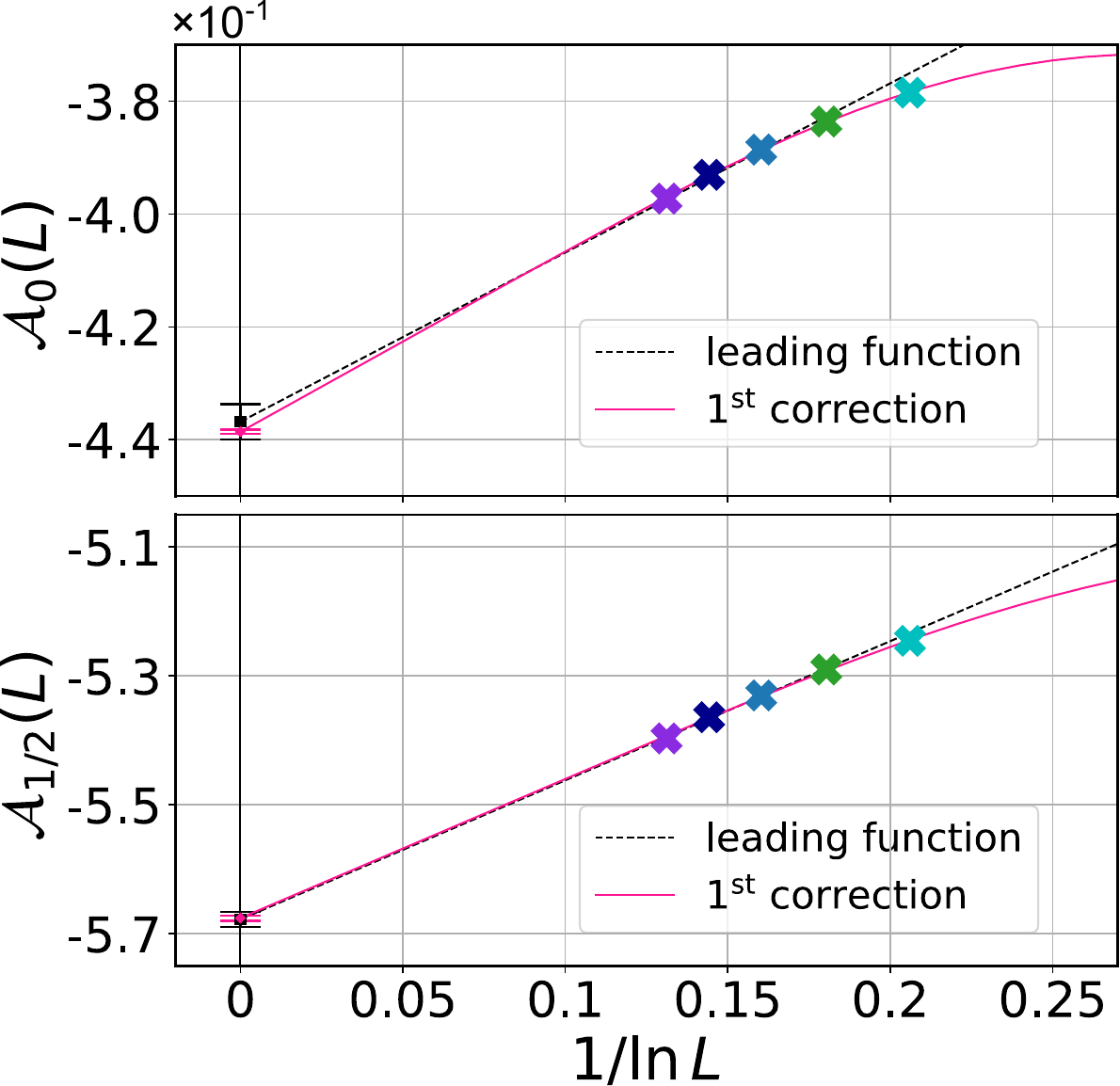}
\centering
\vspace{-4pt}
\caption{Legendre-transform estimate of $\alpha_q$ from $\mathcal{A}_q(L)$ defined in Eq.~(\ref{eq:appD_A_def}). Dashed/solid curves are finite-size extrapolation forms from Eq.~(\ref{eq:appD_A_L}).}
\label{fig:appD_alpha}
\end{figure}

This Appendix summarizes (i) how the parabolic form $\tau_{\max}(q)\approx-\gamma q^2+\nu q$ is extracted from finite-size scaling of extreme moments, (ii) how $\nu$ and $\gamma$ follow from logarithmic moments, and (iii) how the associated Legendre variables $\alpha_q$ and the spectrum $f_{\max}(\alpha)$ are obtained and extrapolated in the presence of irrelevant corrections.

\paragraph{Extreme-moment scaling and parabolic fit.}
We define the extreme-moment exponent by
\begin{equation}\label{eq:appD_tau_q}
\mathbb{E}\!\left[|\psi|_{\max}^{2q}\right]\sim L^{-d\,\tau_{\max}(q)}\qquad (L\to\infty),
\end{equation}
and in the accessible window $|q|\lesssim 1$ we find the parabolic form
\begin{equation}\label{eq:appD_parabolic}
\tau_{\max}(q)\approx -\gamma q^2+\nu q .
\end{equation}

\paragraph{Extracting $\nu$ and $\gamma$ from logarithmic moments.}
Differentiating $\ln \mathbb{E}\!\left[(|\psi|_{\max})^{2q}\right]$ at $q=0$ gives
\begin{equation}\label{eq:appD_nu}
\mathbb{E}\!\left[\ln |\psi|_{\max}^2\right]\sim -d\,\nu\,\ln L ,
\end{equation}
i.e.\ for $d=2$ equivalently $\mathbb{E}[\ln |\psi|_{\max}]\sim -\nu\ln L$.
Evaluating Eq.~(\ref{eq:appD_tau_q}) at $q=\tfrac12$ (so that $\mathbb{E}[(|\psi|_{\max})^{2q}]=\mathbb{E}[|\psi|_{\max}]$) yields the Jensen gap
\begin{equation}\label{eq:appD_gamma}
J_{\ln}(|\psi|_{\max})\coloneqq \ln \mathbb{E}[|\psi|_{\max}]-\mathbb{E}[\ln |\psi|_{\max}]
\sim \frac{d\,\gamma}{4}\,\ln L .
\end{equation}
Linear fits of the right-hand sides of Eqs.~(\ref{eq:appD_nu}) and (\ref{eq:appD_gamma}) versus $\ln L$ give the values quoted in the main text, $\nu=-0.430\pm0.002$ and $\gamma=0.137\pm0.0007$.

\subsection*{Legendre transformation and $\alpha_q$}
To match the main-text convention and keep the Legendre transform in standard form, we define the large-deviation variable $\alpha$ by
\begin{equation}\label{eq:appD_alpha_def}
|\psi|_{\max}^2 \coloneqq L^{-d\,\alpha}
\quad\Longleftrightarrow\quad
\alpha=-\frac{\ln |\psi|_{\max}^{2}}{d\,\ln L}.
\end{equation}
The corresponding spectrum $f_{\max}(\alpha)$ is defined through $\mathrm{PDF}_L(\alpha)\asymp L^{d f_{\max}(\alpha)}$.
A saddle-point evaluation yields the usual Legendre relations
\begin{align}
f_{\max}(\alpha)&= \alpha q-\tau_{\max}(q),\label{eq:appD_Le_Trans}\\
\alpha&= \partial_q\tau_{\max}(q),\label{eq:appD_Le_alpha}
\end{align}
with saddle condition $q=\partial_\alpha f_{\max}(\alpha)$.
We denote by $\alpha_q$ the Legendre saddle associated with moment $q$,
\begin{equation}\label{eq:appD_alpha_q}
\alpha_q\coloneqq \partial_q\tau_{\max}(q).
\end{equation}
Using Eq.~(\ref{eq:appD_parabolic}) gives $\alpha_q=-2\gamma q+\nu$, and eliminating $q$ in Eq.~(\ref{eq:appD_Le_Trans}) yields
\begin{equation}\label{eq:appD_f_alpha}
f_{\max}(\alpha)=-\frac{(\alpha-\nu)^2}{4\gamma}.
\end{equation}

\subsection*{Finite-size extrapolation of $\alpha_q$}
Following Ref.~\cite{Obu2012}, we include irrelevant corrections through
\begin{equation}\label{eq:appD_LF}
\mathbb{E}\!\left[(|\psi|_{\max})^{2q}\right]=L^{-d\,\tau_{\max}(q)}F(q,L^{-y}),
\end{equation}
with $y>0$ and $F$ expandable in powers of $L^{-y}$.
Define the finite-size estimator
\begin{equation}\label{eq:appD_A_def}
\begin{aligned}
\mathcal{A}_q(L)
&\coloneqq -\frac{1}{d\,\ln L}\frac{\mathrm{d}}{\mathrm{d}q}
\ln \mathbb{E}\!\left[(|\psi|_{\max})^{2q}\right] \\
&= -\frac{\mathbb{E}\!\left[(|\psi|_{\max})^{2q}\ln |\psi|_{\max}^2\right]}
{d\,\ln L\,\mathbb{E}\!\left[(|\psi|_{\max})^{2q}\right]} .
\end{aligned}
\end{equation}
Inserting Eq.~(\ref{eq:appD_LF}) gives
\begin{equation}\label{eq:appD_A_split}
\begin{aligned}
\mathcal{A}_q(L)
&=\partial_q\tau_{\max}(q)-\frac{1}{d\,\ln L}\,\partial_q\ln F(q,L^{-y})\\
&=\alpha_q-\frac{1}{d\,\ln L}\,\partial_q\ln F(q,L^{-y}) .
\end{aligned}
\end{equation}
Expanding $\partial_q\ln F$ in powers of $L^{-y}$ yields the extrapolation form
\begin{equation}\label{eq:appD_A_L}
\mathcal{A}_q(L)=\alpha_q+\frac{b_q}{\ln L}\left(1+c_qL^{-y}+d_qL^{-2y}+\cdots\right),
\end{equation}
which is used for the fits in Fig.~\ref{fig:appD_alpha}. The coefficients $c_q$ and higher-order correction amplitudes are nonuniversal and serve to control extrapolation quality \cite{Obu2012}.

Hence $\lim_{L\to\infty}\mathcal{A}_q(L)=\alpha_q$. The extrapolation of $\mathcal{A}_q(L)$ is shown in Fig.~\ref{fig:appD_alpha}. Using the form above truncated at the $L^{-y}$ correction, we obtain $\alpha_0=-0.4386\pm0.0004$ with $y=0.936\pm0.15$ and $b_q=0.31\pm0.01$, and $\alpha_{1/2}=-0.5676\pm0.0004$. The coefficients $c_q$ and higher-order correction amplitudes are nonuniversal and less relevant for the present analysis \cite{Obu2012}.

Although these extrapolations are individually of high quality ($\bar{R}^2\gtrsim 0.999$), the values of $\alpha_0$ and $\alpha_0-\alpha_{1/2}$ differ from $\nu$ and $\gamma$ by about $0.005$. We attribute this residual mismatch to finite-size limitations; the largest efficiently accessible system size in our simulations is $L=2048$.

\section{Appendix D: Full-state multifractal benchmark}
\label{app:C}

\renewcommand{\thefigure}{D\arabic{figure}}
\setcounter{figure}{0}
\renewcommand{\theequation}{D\arabic{equation}}
\setcounter{equation}{0}

\begin{figure}[t]
\includegraphics[width=\columnwidth]{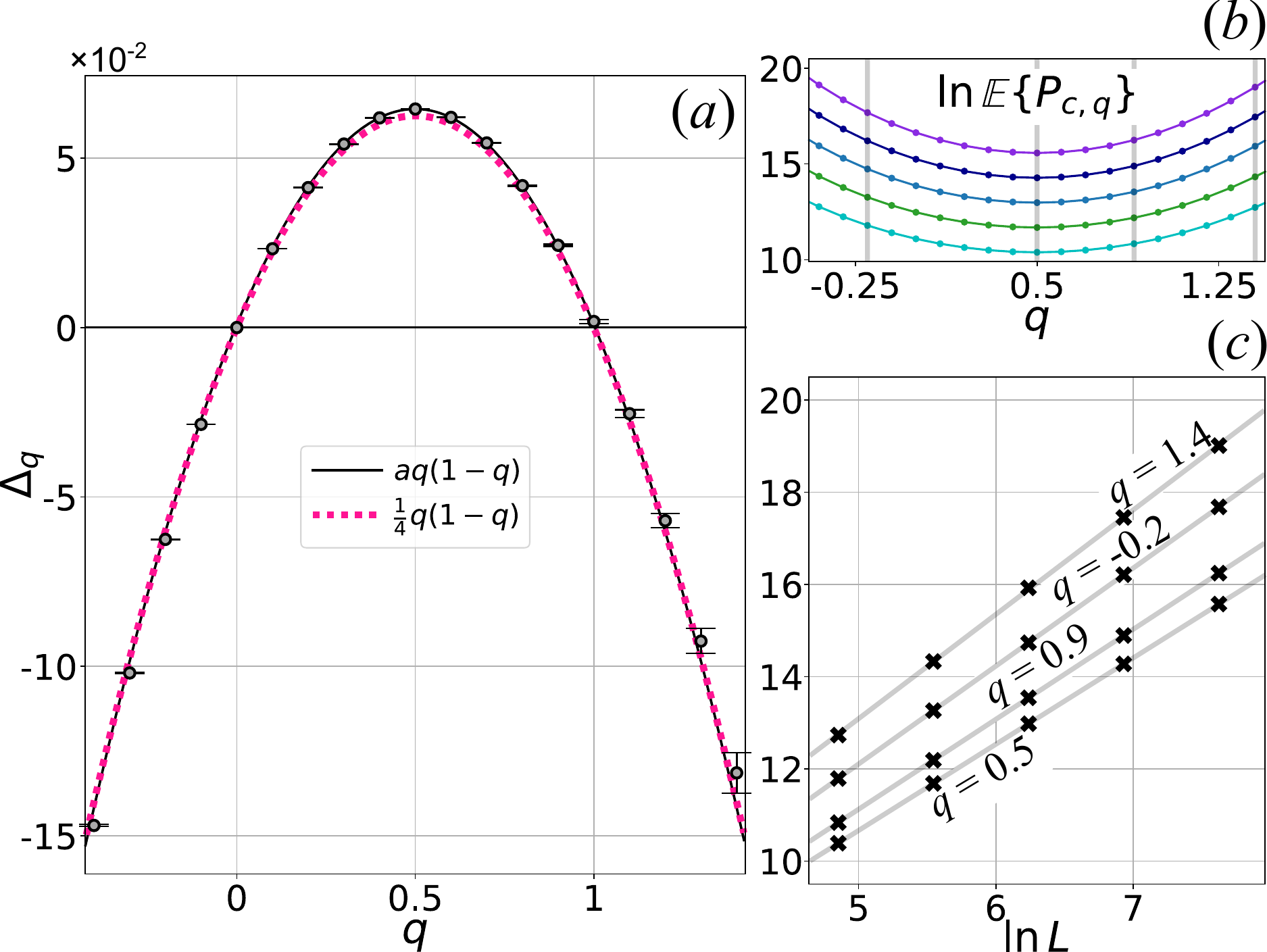}
\centering
\caption{Multifractal exponent for the full stationary state. (a) Measured $\Delta_q$ compared with the parabolic form $\frac{1}{4}q(1-q)$ (dashed curve) and an unconstrained best parabolic fit (solid curve). (b) $\ln \mathbb{E}\{P_{c,q}\}$; data from bottom to top correspond to $L=64,128,\ldots,2048$. (c) For each $q$, the value of $\Delta_q$ is extracted from the slope of $\ln \mathbb{E}\{P_{c,q}\}$ versus $\ln L$.}
\label{fig:appC_Deltaq}
\end{figure}

As a numerical sanity check, we extract the standard multifractal exponent $\Delta_q$ from the stationary state $|\varPsi\rangle=\sum_l\psi_l|l\rangle$ and verify consistency with the established IQH literature. For $l\neq c$, the two-point moment in the open point-contact geometry is expected to decay algebraically \cite{Bon2014,KleZir2001},
\begin{equation}\label{eq:appC_two_point}
\mathbb{E}\!\left[|\psi_{l\neq c}|^{2q}\right]\propto |l-c|^{-2\Delta_q}.
\end{equation}
The exponent $\Delta_q$ coincides with the multifractal exponent of the ordinary closed network and obeys the symmetry $\Delta_q=\Delta_{1-q}$ \cite{Mir2006,Eve2008}. While $\Delta_q$ is approximately parabolic in the accessible range, high-precision studies report small but significant deviations from perfect parabolicity \cite{Obu2012,Bab2023,EversMildenbergerMirlin2008}.

To probe the integrated scaling, we define $P_{c,q}\equiv \sum_{l\neq c}|\psi_l|^{2q}$ (omitting the fixed contact amplitude). Using Eq.~(\ref{eq:appC_two_point}) gives $\mathbb{E}[P_{c,q}]\propto \sum_{l\neq c}|l-c|^{-2\Delta_q}\sim L^{-2\bar{\Delta}_q}$, and in the continuum approximation one finds $\sum_{l\neq c}|l-c|^{-2\Delta_q}\sim \int_1^L x^{-2\Delta_q}\,2\pi x\,dx\sim L^{-2\Delta_q+2}$ (in the regime where the large-distance contribution dominates), implying $\Delta_q=\bar{\Delta}_q+1$ for large $L$. Figure~\ref{fig:appC_Deltaq} shows that our extracted $\Delta_q$ agrees with the established trends reported in Refs.~\cite{Eve2001,Eve2008,Obu2012,Zir2019}.

\section{Appendix E: Global gain and normalized maxima}
\label{app:E}

\renewcommand{\thefigure}{E\arabic{figure}}
\setcounter{figure}{0}
\renewcommand{\theequation}{E\arabic{equation}}
\setcounter{equation}{0}

\begin{figure}[t]
\includegraphics[width=0.9\columnwidth]{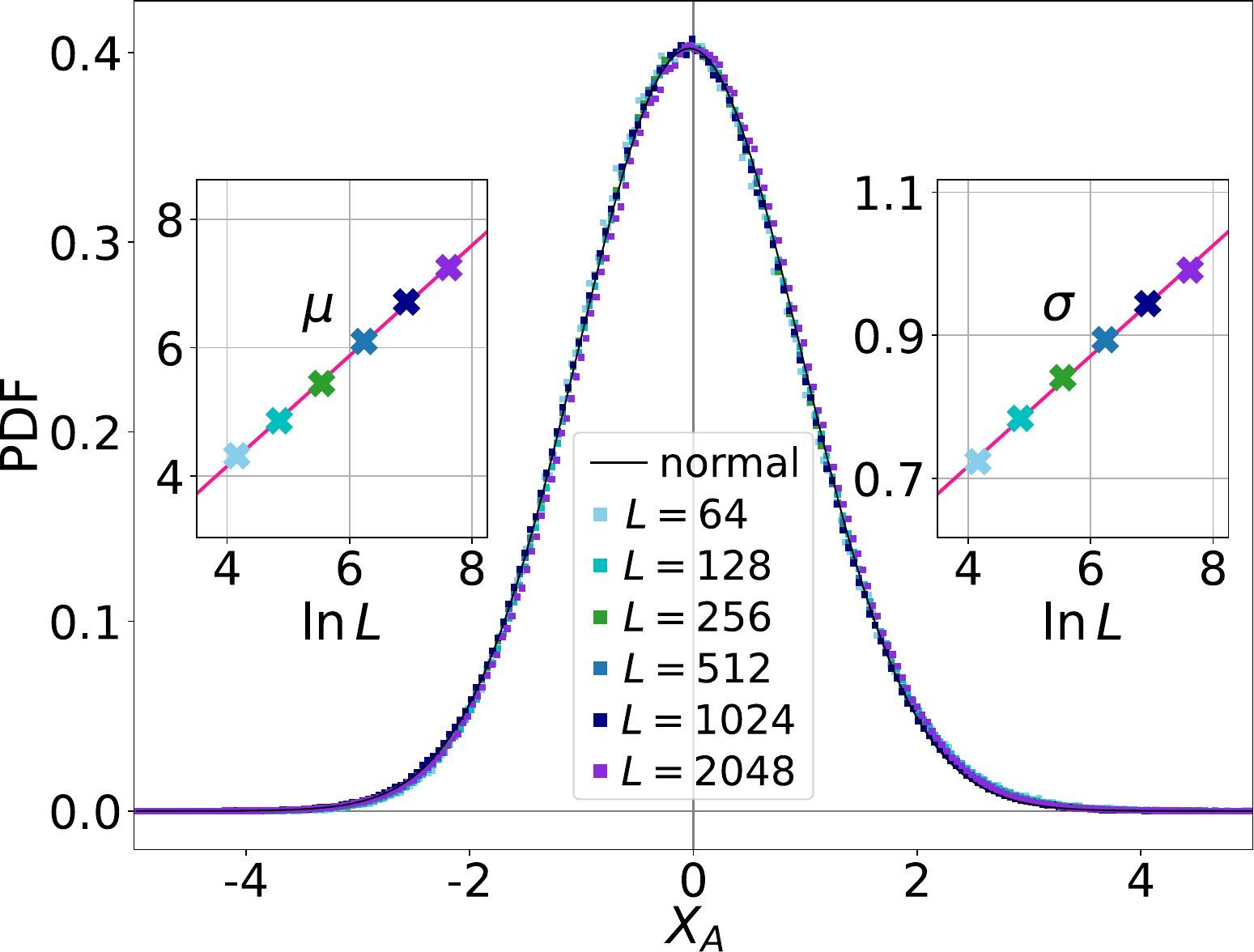}
\centering
\caption{Probability densities of the global gain. The curves are shown using the standardized variable $X_A=(\ln A-\mu)/\sigma$, where $\mu$ and $\sigma$ are the mode and standard deviation of $\mathrm{PDF}(\ln A)$, respectively. The distributions are close to normal, and both $\mu$ and $\sigma$ vary approximately linearly with $\ln L$.}
\label{fig:appE_lnA}
\end{figure}

\begin{figure}[b]
\includegraphics[width=0.9\columnwidth]{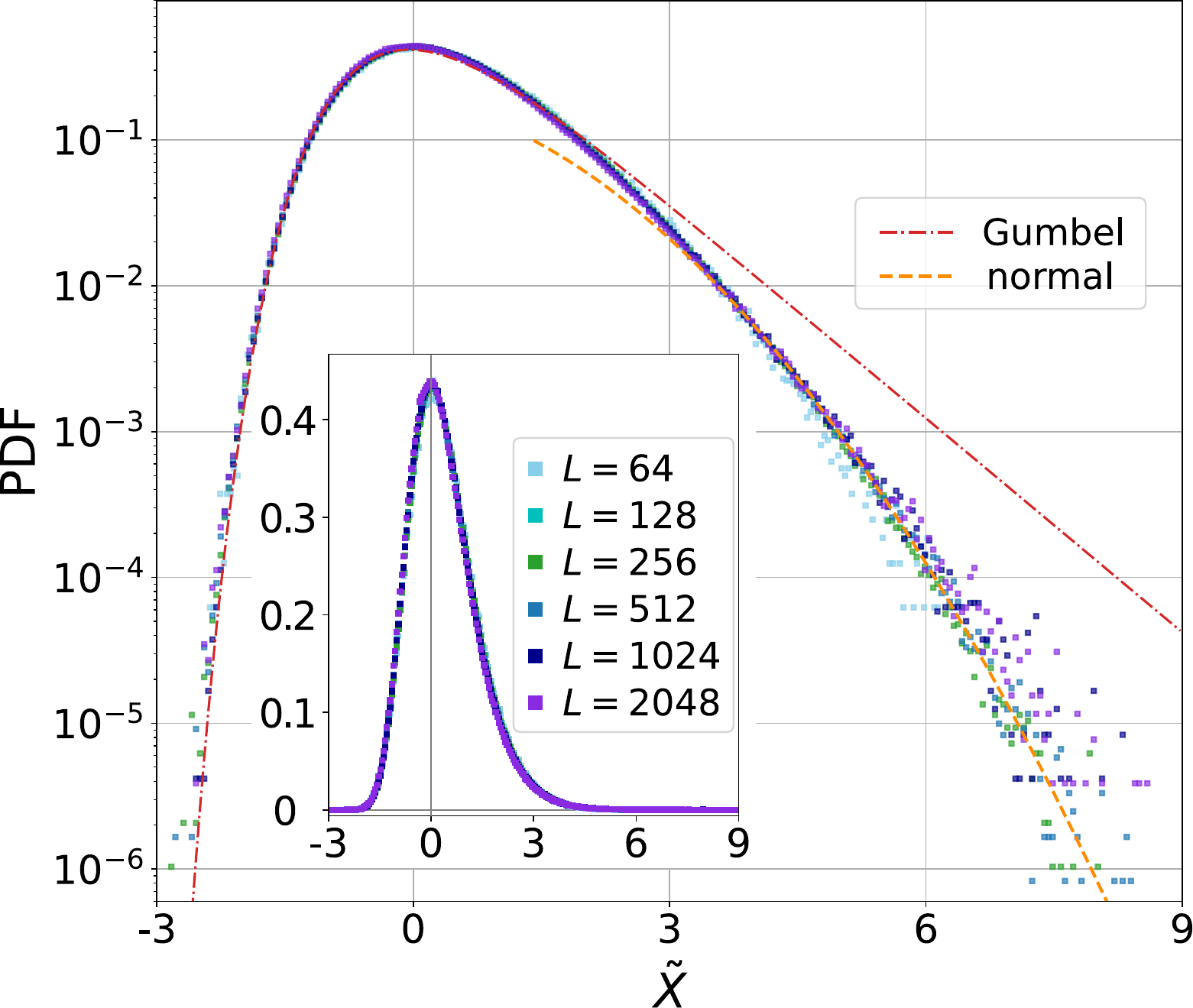}
\centering
\caption{Probability densities of the normalized maxima. The standardized variable is $\tilde{X}=(\ln |\tilde\psi|_{\max}-\mu)/\sigma$, where $\mu$ and $\sigma$ are the mode and standard deviation of $\mathrm{PDF}(\ln |\tilde\psi|_{\max})$, respectively. The distribution exhibits a compound form: a Gumbel-like regime at smaller values and a crossover to a Gaussian decay in the right tail of $\ln|\tilde\psi|_{\max}$.}
\label{fig:appE_lnNorm}
\end{figure}

This Appendix records the statistics of the sample-dependent gain factor and summarizes how it shapes the raw and normalized extreme-value observables. We define the global gain by
$A\coloneqq \left(\sum_{l\neq c} |\psi_l|^2\right)^{1/2}$.
 Figure~\ref{fig:appE_lnA} shows $\mathrm{PDF}(\ln A)$, which is close to Gaussian after standardization; moreover, both the mode $\mu$ and the width $\sigma$ vary approximately linearly with $\ln L$ (fit quality $\bar{R}^2\gtrsim 0.999$). Since $|\psi|_{\max}=A\,|\tilde\psi|_{\max}$ by definition, $\ln|\psi|_{\max}=\ln A+\ln|\tilde\psi|_{\max}$; therefore, if $\ln A$ has mean and variance proportional to $\ln L$ and higher cumulants are subleading over the accessible sizes, then $\ln|\psi|_{\max}$ is naturally close to Gaussian in the same regime and the raw extreme moments acquire an approximately quadratic dependence on $q$, producing the parabolic $\tau_{\max}(q)$ reported in the main text.

Figure~\ref{fig:appE_lnNorm} shows $\mathrm{PDF}(\ln|\tilde\psi|_{\max})$ for the gain-normalized maximum $|\tilde\psi|_{\max}\equiv |\psi|_{\max}/A$. In contrast to the raw case, the normalized distribution is not close to a single log-normal form: it displays a Gumbel-like regime at smaller $\ln|\tilde\psi|_{\max}$ and a crossover to an approximately Gaussian decay in the right tail of $\ln|\tilde\psi|_{\max}$. Under the same centering/scaling protocol across sizes, we do not observe a collapse consistent with a single-parameter GEV form over the accessible range, supporting the intrinsic, correlation-dominated extreme sector discussed in the main text. Finally, note that even an apparent Gumbel fit for $\ln|\tilde\psi|_{\max}$ would correspond to a \emph{log-Gumbel} law for $|\tilde\psi|_{\max}$ and would not by itself establish standard GEV statistics for the maximum in the original variable.

\end{document}